\documentclass[hyper]{JHEP} 

\usepackage{epsfig}

\newcommand\fverb{\setbox\pippobox=\hbox\bgroup\verb}

\newcommand\fverbdo{\egroup\medskip\noindent%

            \fbox{\unhbox\pippobox}\ }

\newcommand\fverbit{\egroup\item[\fbox{\unhbox\pippobox}]}

\newbox\pippobox

\title{Lagrange Multiplier Modified  Ho\v{r}ava-Lifshitz Gravity}
\author{J. Kluso\v{n}\\
Department of
Theoretical Physics and Astrophysics\\
Faculty of Science, Masaryk University\\
Kotl\'{a}\v{r}sk\'{a} 2, 611 37, Brno\\
Czech Republic\\
E-mail: \email{klu@physics.muni.cz}}
\preprint{}

 \abstract{We consider RFDiff invariant
 Ho\v{r}ava-Lifshitz gravity action
 with additional Lagrange multiplier
 term that is a function of scalar
 curvature. We find its Hamiltonian
 formulation and we  show that
 the constraint structure implies
 the same number of physical degrees
 of freedom as in General Relativity.}

\def\be{\begin{equation}}

\def\ee{\end{equation}}

\def\bea{\begin{eqnarray}}

\def\eea{\end{eqnarray}}

\def\tK{\tilde{K}}

\def\mH{\mathcal{H}}

\def\bz{\mathbf{z}}

\def\bx{\mathbf{x}}
\def\by{\mathbf{y}}

\newcommand{\hg}{\hat{g}}

\newcommand{\mA}{\mathcal{A}}

\newcommand{\mG}{\mathcal{G}}

\def\mV{\mathcal{V}}

\newcommand{\bT}{\mathbf{T}}

\def\pb #1{\left\{#1\right\}}

\begin{document}
\section{Introduction and Summary}\label{first}
In 2009 Petr Ho\v{r}ava formulated new
proposal of quantum theory of gravity
(now known as
Ho\v{r}ava-Lifshitz gravity (HL
gravity)
that is power counting renormalizable
\cite{Horava:2009uw,Horava:2008ih,Horava:2008jf}
that is also expected that it
reduces do General Relativity in the
infrared (IR) limit
\footnote{For review and
extensive list of references, see
\cite{Horava:2011gd,Padilla:2010ge,Mukohyama:2010xz,Weinfurtner:2010hz}.}.
The HL gravity is based on
 an idea that
the Lorentz symmetry is restored in IR
limit of given theory while it is absent
in its  high energy regime. For that reason
  Ho\v{r}ava considered
systems whose scaling at short
distances exhibits a strong anisotropy
between space and time,
\begin{equation}
\bx' =l \bx \ , \quad t' =l^{z} t \ .
\end{equation}
In $(D+1)$ dimensional space-time in
order to have power counting
renormalizable theory  requires that
$z\geq D$. It turns out however that
the symmetry group of given theory is
reduced from the full diffeomorphism
invariance of General Relativity  to
the foliation preserving diffeomorphism
\begin{equation}\label{fpdi}
x'^i=x^i+\zeta^ i(t,\bx) \ , \quad
t'=t+f(t) \ .
\end{equation}
Due to the fact that the diffeomorphism
is restricted (\ref{fpdi})  one more
degree of freedom appears that is a
spin$-0$ graviton. It turns out that
the existence of this mode could be
dangerous since it has to decouple
 in the IR regime, in order to
be consistent with observations.
Unfortunately, it seems  that this
might not be the case. It was shown
that the spin-0 mode is not stable in
the original version of the HL theory
[1] as well as in the Sotiriou, Visser
and Weinfurtner (SVW) generalization
\cite{Sotiriou:2009bx}. Note that in
both of these two versions, it was all
assumed the projectability condition
that means that the lapse function $N$
depends on $t$ only. This presumption
has a fundamental consequence for the
formulation of the theory since there
is no local form of the Hamiltonian
constraint but only the global one.

On the other hand we can consider  the
second version of HL gravity where the
projectability condition is not imposed
so that $N=N(\bx,t)$ \footnote{For
another proposal of renormalizable
theory of gravity, see
\cite{Nojiri:2009th,Nojiri:2010tv}.}.
This form of HL gravity was extensively
studied in
\cite{Blas:2009yd,Blas:2009qj,Blas:2009ck,Blas:2010hb,Li:2009bg,Blas:2010hb,
Kluson:2010xx,Kluson:2010nf,Bellorin:2010te,Bellorin:2010je,Kobakhidze:2009zr,Pons:2010ke}.
It was shown  in \cite{Blas:2010hb}
that  so called healthy extended
version of given theory could really be
an interesting candidate for the
quantum theory of gravity without
ghosts and without strong coupling
problem despite its unusual Hamiltonian
structure
\cite{Kluson:2010xx,Kluson:2010nf}.

Recently Ho\v{r}ava and Malby-Thompson
in \cite{Horava:2010zj}
 proposed very interesting way how
to eliminate the spin-0 graviton in the context
of the projectable version of HL
gravity. Their construction is
based on an
 extension of
the foliation preserving diffeomorphism in
such a way that the theory is invariant
under additional
 local $U(1)$ symmetry. The
resulting theory is known  as
non-relativistic  covariant theory of
gravity \footnote{This theory was also
studied in
\cite{Greenwald:2010fp,Alexandre:2010cb,Wang:2010wi,Huang:2010ay,Kluson:2010za,Kluson:2010zn}.}.
It was shown in
\cite{Horava:2010zj,daSilva:2010bm}
that the presence of this new symmetry
implies that the spin-0 graviton
becomes non-propagating and the
spectrum of the linear fluctuations
around the background solution
coincides with the fluctuation spectrum
of General Relativity.

In this paper we present another
version of HL gravity
 with the correct
number of physical degrees of freedom.
Our model is based on the formulation
of the HL gravity with reduced symmetry
group known as
\emph{restricted-foliation-preserving
Diff} (RFDiff)
 HL gravity
\cite{Blas:2010hb,Kluson:2010na}. This
is the theory that is invariant under
following symmetries
\begin{equation}\label{rfdtr}
t'=t+\delta t \ , \delta
t=\mathrm{const} \ , \quad x'^i=
x^i+\zeta^i(\bx,t) \ .
\end{equation}
The characteristic property of given
theory is the absence of the Hamiltonian
constraint  \cite{Kluson:2010na} either
global or local. Note that the meaning
of the global Hamiltonian constraint
is not completely clear
\cite{Horava:2010zj} so that formulation
of the HL gravity without the lapse function
could be an interesting possibility how
to eliminate this problem.
Our construction is based on the idea of
modification of RFDiff HL  action
that respects all symmetries of the theory
however which changes the constraint
structure of given theory.
Remarkably this goal can be achieved when
we include into the action additional
term which is a function of the scalar
curvature and it is multiplied by
Lagrange multiplier.
 Then we perform the
 Hamiltonian analysis of
given  system and we show
 that the number of
physical degrees of freedom coincides
with the physical  number of degrees freedom of
General Relativity. This
fact implies that
 dangerous scalar graviton is
eliminated  even if there is no additional
 gauge symmetry
\footnote{We would like to stress that
the Lagrange multiplier modification of
RFDiff HL gravity presented in this paper
can be easily extended to projectable
version of HL gravity as well.}.
This remarkable result suggests that
 Lagrange multiplier modified
RFDiff HL gravity  is an interesting
example of the  power counting
renormalizable theory of gravity with
restricted symmetry group where however
the scalar graviton is eliminated.
On the other hand the fact that this is theory
with the second class constraints
 makes the
deeper analysis rather obscure. In fact,
it is not clear how to solve the second
class constraints for the physical degrees of freedom.
Further, due to the fact that the Poisson
bracket between second class constraints
depends on the phase space variables implies
that the  symplectic
structure  on the reduced phase space
 defined by corresponding Dirac brackets
depends on phase space variables which makes
further analysis of given theory very difficult.
In fact, the conventional method for
covariant quantization of a theory with
second class constraints is to go
 over to an equivalent formulation
where second class constraints
 are replaced by the first class ones in
one or another way. For example,
implementing the abelian conversion of
the second class constraints
 \cite{Batalin:1991jm}
we can formulate given theory as the
theory with the  first class constraints.
Explicitly, by introducing additional
variables $\Phi$ called conversion
variables we can extend  second-class constraints by
$\Phi$ dependent terms such
that the extended constraints become first class.
However we can expect that given procedure
will be purely formal due to the fact that
the Poisson bracket between the second
class constraints depends on the phase
space variables in complicated way
so that the extended Hamiltonian and constraints
will contain infinite many terms.

We would like to stress that
  the Lagrange multiplier modified gravities
were studied previously in
\cite{Capozziello:2010uv}, see also
\cite{Kluson:2010af,Cai:2010zma}.
However due to the fact that given
theories are invariant under full
diffeomorphism   it is only
possible to add to the action additional terms
that are functions of the space-time
curvature only. As a result the Hamiltonian
structure of given theory is in agreement with
basic principles of geometrodynamics
\cite{Isham:1984sb,Isham:1984rz,Hojman:1976vp}.
In other words, the Hamiltonian structure
of Lagrange multiplier modified $F(R)$ gravities
is the same as the structure of the original
$F(R)$ gravity. We can generalize this
construction and
 consider
 the Lagrange multiplier
modified $F(\tilde{R})$ HL gravity
\cite{Carloni:2010nx,Chaichian:2010yi},
for review see \cite{Nojiri:2010wj}.
Even if the resulting theory can be
interesting in its own it cannot solve
the scalar graviton problem of HL
gravity due to the presence of
additional scalar modes that are
general property of all $F(R)$ theories
of gravity. More precisely, the
Hamiltonian structure of Lagrange
multiplier modified $F(\tilde{R})$ HL
gravity is the same as the Hamiltonian
structure of $F(\tilde{R})$ HL gravity
coupled to scalar field with specific
form of the action. As a consequence
the resulting constraints are not
sufficient to eliminate the scalar
graviton. We should however stress that
we could consider yet another form of
the Lagrange modified $F(\tilde{R})$ HL
gravities where we add additional term
that is function of the scalar
curvature $R$ instead of $\tilde{R}$.
It is easy to see that the presence of
the additional constraint is sufficient
for the elimination of the scalar
graviton.

Let us outline our results and suggest
possible extension of this work.
We consider Lagrange multiplier modified
RFDiff invariant HL gravity and we argue
that the number of physical degrees of freedom
coincides with the number of degrees
of freedom of General Relativity. As a consequence
the scalar graviton can be eliminated
in the fluctuation spectrum of given theory.
If we combine this result with the well known
fact that HL gravity is power counting
renormalizable theory we derive an intriguing
formulation of the theory of gravity that has
correct number of physical degrees of freedom
and which is
potentially power counting renormalizable.
Of course there is still the problematic fact
that this is the theory  with the second
class constraints. The related problem is
that this is the theory with the complicated
 symplectic structure.

Let us suggest possible extensions of given work.
It would be nice to see whether
the Lagrange multiplier mechanism can be implemented
in the structure of
infrared modified gravities
(For review, see \cite{Rubakov:2008nh})
and solve some of their problems.  There
is also an open question  how the low energy
limit of  the Lagrange
multiplier modified RFDiff HL gravity
is related to General Relativity.

This paper is organized as follows. In
the next section we  review basic
properties of  RFDiff  HL
gravity and perform its modification
when we include term multiplied by
Lagrange multiplier into corresponding
action. Then we find its Hamiltonian
formulation and determine constraints
structure.  In section
(\ref{third}) we consider  more general
form of Lagrange multiplier modified
RFDiff invariant HL gravities and
analyze their properties.

\section{Ho\v{r}ava-Lifshitz Gravity with
Lagrange Multiplier }\label{second} We
begin  this section with review
of basic facts  needed
for the formulation of RFDiff invariant
HL gravity. This is the well know $D+1$ formalism
that is the fundamental ingredient of the Hamiltonian
formalism of any theory of gravity
\footnote{For recent review,
see \cite{Gourgoulhon:2007ue}.}.

Let us consider $D+1$ dimensional
manifold $\mathcal{M}$ with the
coordinates $x^\mu \ , \mu=0,\dots,D$
and where $x^\mu=(t,\bx) \ ,
\bx=(x^1,\dots,x^D)$. We presume that
this space-time is endowed with the
metric $\hat{g}_{\mu\nu}(x^\rho)$ with
signature $(-,+,\dots,+)$. Suppose that
$ \mathcal{M}$ can be foliated by a
family of space-like surfaces
$\Sigma_t$ defined by $t=x^0$. Let
$g_{ij}, i,j=1,\dots,D$ denotes the
metric on $\Sigma_t$ with inverse
$g^{ij}$ so that $g_{ij}g^{jk}=
\delta_i^k$. We further introduce the operator
$\nabla_i$ that is covariant derivative
defined with the metric $g_{ij}$.
 We  introduce  the
future-pointing unit normal vector
$n^\mu$ to the surface $\Sigma_t$. In
ADM variables we have
$n^0=\sqrt{-\hat{g}^{00}},
n^i=-\hat{g}^{0i}/\sqrt{-\hat{g}^{
00}}$. We also define  the lapse
function $N=1/\sqrt{-\hat{g}^{00}}$ and
the shift function
$N^i=-\hat{g}^{0i}/\hat{g}^{00}$. In
terms of these variables we write the
components of the metric
$\hat{g}_{\mu\nu}$ as
\begin{eqnarray}
\hat{g}_{00}=-N^2+N_i g^{ij}N_j \ ,
\quad \hat{g}_{0i}=N_i \ , \quad
\hat{g}_{ij}=g_{ij} \ ,
\nonumber \\
\hat{g}^{00}=-\frac{1}{N^2} \ , \quad
\hat{g}^{0i}=\frac{N^i}{N^2} \ , \quad
\hat{g}^{ij}=g^{ij}-\frac{N^i N^j}{N^2}
\ .
\nonumber \\
\end{eqnarray}
 RFDiff invariant Ho\v{r}ava-Lifshitz
 gravity was introduced in
\cite{Blas:2010hb} and further studied in
\cite{Kluson:2010na}. This is the version
of the Ho\v{r}ava-Lifshitz gravity that
is  not invariant under
 foliation preserving diffeomorphism
 but only under reduced set of diffeomorphism
\begin{equation}\label{RFDtr}
t'=t+\delta t \ , \quad  \delta
t=\mathrm{const} \ , \quad  x'^i=
x^i+\xi^ i(t,\bx)
\end{equation}
The simplest form of RFDiff invariant
Ho\v{r}ava-Lifshitz gravity takes the
form \cite{Kluson:2010na}
\begin{equation}\label{RFDiffaction}
S=\frac{1}{\kappa^2} \int dt d^D\bx
\sqrt{g}(\tK_{ij}
\mG^{ijkl}\tK_{kl}-\mV(g)) \ ,
\end{equation}
where we introduced modified extrinsic
curvature
\begin{equation}
\tK_{ij}=\frac{1}{2}(\partial_t g_{ij}
-\nabla_i N_j-\nabla_j N_i) \
\end{equation}
that differs from the standard extrinsic
curvature by absence of the lapse $N(t)$.
Further the generalized De Witt metric
$\mG^{ijkl}$  is defined as
\begin{equation}
\mG^{ijkl}=\frac{1}{2}(g^{ik}g^{jl}+
g^{il}g^{jk})-\lambda g^{ij}g^{kl} \ ,
\end{equation}
where $\lambda$ is a real constant that in
case of General Relativity is equal to one.
Finally $\mV(g)$ is general function
of $g_{ij}$ and its covariant derivative.

We would like to stress that   the
action (\ref{RFDiffaction}) differs
from the projectable version of HL
gravity by absence of the lapse
$N=N(t)$. It is clear that we  could
consider more general form of RFDiff HL
theory where the time and space partial
derivatives of $N$ are included
\cite{Blas:2010hb}. Due to the fact
that $N$ behaves as scalar under
(\ref{RFDtr}) we can interpret this
theory as a coupled system of the
RFDiff HL gravity (\ref{RFDiffaction})
with  the scalar field. Then for
simplicity we restrict ourselves to the
action (\ref{RFDiffaction}) keeping in
mind that it can be easily generalized.


It was shown in \cite{Kluson:2010zn}
that this action can be
extended to be invariant under $U(1)$
transformation, following very nice
construction given in \cite{Horava:2010zj}.
 Hamiltonian analysis of given
theory shows that the number of physical
degrees of freedom coincides with the
number of physical degrees of freedom
of General Relativity. Now we show
 that the same
result can be derived with the  minimal
extension of RFDiff  HL gravity
when we add to the original
RFDiff HL action following
term
\begin{equation}\label{Snuk}
S_{l.m.}= \frac{1}{\kappa^2}\int dt
d^D\bx  \sqrt{g} \mG(R)\mA \ ,
\end{equation}
where $\mG$ is  function of
$D-$dimensional curvature $R$
\footnote{ $\mG(R)=R-\Omega$ in
\cite{Horava:2010zj}.} and where $\mA$
is  Lagrange multiplier that
transforms as scalar
\begin{equation}
\mA'(t',\bx')=
\mA(t,\bx) \
\end{equation}
under (\ref{RFDtr}). Then since $R$ and
$ d^D\bx \sqrt{g}$ are manifestly
invariant under (\ref{RFDtr})  we
immediately obtain that (\ref{Snuk}) is
invariant under  (\ref{RFDtr}).

In summary we consider following action for
Lagrange multiplier modified RFDiff
HL gravity
\begin{equation}\label{RFDaction}
S=\frac{1}{\kappa^2} \int dt
d^D\bx \sqrt{g}(\tK_{ij}
\mG^{ijkl}\tK_{kl}-\mV(g)+\mG(R)\mA)
\ .
\end{equation}
Our goal is to perform the Hamiltonian
analysis of given theory.
 From (\ref{RFDaction}) we find the conjugate momenta
\begin{eqnarray}\label{piK}
\pi^{ij}&=& \frac{1}{\kappa^2}\sqrt{g}\mG^{ijkl}\tK_{kl}
\ ,  \quad p^i
\approx 0 \ , \quad
p_\mA \approx 0 \  \quad
\nonumber \\
\end{eqnarray}
that imply the $D+1$
primary constraints
\begin{equation}
p_i(\bx)\approx 0 \ ,  \quad
p_\mA(\bx)\approx 0 \ .
\end{equation}
Further, using (\ref{piK}) we easily
 find the
 Hamiltonian with primary constraints
 included
\begin{eqnarray}\label{HRFD}
H&=&\int d^D\bx (\mH_T+N^i\mH_i+
v^\mA p_\mA+v_i p^ i)-
\frac{1}{\kappa^2}
\int d^D\bx \sqrt{g}\mG(R)
\mA \ ,
\nonumber \\
\end{eqnarray}
where
\begin{eqnarray}\label{mHTAr1}
\mH_T&=&\frac{\kappa^2}{\sqrt{g}}\pi^{ij}\mG_{ijkl}\pi^{kl}
-\frac{1}{\kappa^2}
\sqrt{g}\mV(g) \ ,
\nonumber \\
\mH_i&=&-2g_{il} \nabla_k \pi^{kl} \ . \nonumber \\
\end{eqnarray}
Following standard analysis of the
constraint systems
\cite{Henneaux:1992ig,Govaerts:2002fq,Govaerts:1991gd}
 we  demand
that the primary constraints are
preserved during the time evolution of
the system. Explicitly
\begin{eqnarray}
\partial_t p_\mA &=&\pb{p_\mA,H}=
-\frac{1}{\kappa^2}
\sqrt{g}\mG(R)\equiv - \Phi_1 \approx 0
\ ,
\nonumber \\
\partial_t p_i&=&\pb{p_i,H}=
-\mH_i\approx 0
 \    \nonumber \\
\end{eqnarray}
so that the requirement of the
preservation of the primary constraints
implies the secondary ones
$\Phi_1\approx 0 , \mH_i\approx 0$. Of
course, now we have to demand that
these constraints are preserved during
the time evolution of the system.
 In case of $\mH_i$ it is
convenient to introduce following
extended smeared form of these
constraints
\begin{equation}
\bT_S(N^i)=
\int d^D\bx N^i(\mH_i+p_\mA \partial_i\mA) \ ,
\end{equation}
where we included the primary
constraint $p_\mA\approx 0$  into the
definition of $\bT_S$. Then it is easy
to see that $\bT_S(N^i)$ is generator
of spatial diffeomorphism. If we
include these secondary constraints
into Hamiltonian we find that the total
Hamiltonian now takes the form
\begin{eqnarray}
H_T&=&\int d^D\bx (\mH_T+
v^{\mA}p_{\mA}+v_ip^i+v^1\Phi_1)+\bT_S(N^i)
\ . \nonumber \\
\end{eqnarray}
Using the fact that the action
is invariant under spatial diffeomorphism
we immediately find that $\bT_S(N^i)$ is
preserved during the time evolution of
the system. The situation is different
in case of the secondary constraint $\Phi_1$.
Using following formulas
\begin{eqnarray}
\pb{R(\bx),\pi^{ij}(\by)}&=&
-R^{ij}(\bx)\delta(\bx-\by)+ \nabla^i
\nabla^j \delta(\bx-\by)-g^{ij}
\nabla_k \nabla^k\delta(\bx-\by) \ ,
\nonumber \\
\nabla^i \nabla^j \mG_{ijkl}
\pi^{kl}&-& g^{ij}\nabla_m\nabla^m
\mG_{ijkl}\pi^{kl} =\nabla_k (\nabla_l
\pi^{kl}) +\frac{1-\lambda}{\lambda
D-1}\nabla_i
\nabla^i \pi \  \nonumber \\
\end{eqnarray}
we find that the time derivative of
$\Phi_1$ is equal to
 \begin{eqnarray}\label{tPhi2}
\partial_t \Phi_1&=&
\pb{\Phi_1,H}\approx -2
\frac{d\mG}{dR} \left(
 R^{ij}
\mG_{ijkl}\pi^{kl} -
\frac{1-\lambda}{\lambda D-1}\nabla_k
\nabla^k \pi\right)=2\frac{d\mG}{dR}
\Phi_2 \ ,
\nonumber \\
\end{eqnarray}
where
\begin{eqnarray}
\Phi_2
=- R_{ij}\pi^{ji}
+\frac{\lambda}{D\lambda-1}R\pi +
\frac{1-\lambda}{\lambda D-1}\nabla_k
\nabla^k \pi \equiv
M_{ij}(g(\bx))\pi^{ji}(\bx)
\   \nonumber \\
\end{eqnarray}
is additional constraint that has to be
imposed in order the constraint
$\Phi_1$ is preserved during the time
evolution of the system. Following
 \cite{Govaerts:2002fq,Govaerts:1991gd}
 we include the constraint $\Phi_2$ into
 definition of the total Hamiltonian so
 that
\begin{eqnarray}
H_T&=&\int d^D\bx (\mH_T+ v^\mA
p_\mA+v_ip^i+ v^1\Phi_1 +v^2
\Phi_2)+\bT_S(N^i) \ , \nonumber \\
\end{eqnarray}
where $v_i,v^{\mA},v^1,v^2$ are
corresponding Lagrange multipliers.

Now we should again check the stability
of all constraints. It is easy to see
that the primary constraints together
with $\bT_S(N^i)$ are preserved while
the   time  evolution of the
constraint $\Phi_1\approx 0$ is equal
to
\begin{eqnarray}\label{parttPhi2}
\partial_t \Phi_1&=&\pb{\Phi_1,H_T}
\approx  \int d^D\bx \left(2
\frac{d\mG}{dR} \Phi_2(\bx) +v^2(\bx)
\pb{\Phi_1,\Phi_2(\bx)}\right)
\approx \nonumber \\
&\approx& \int d^D\bx
v^2(\bx)\pb{\Phi_1,\Phi_2(\bx)}=0 \ .
\nonumber \\
\end{eqnarray}
Since
\begin{eqnarray}\label{DBphi12}
& &\pb{\Phi_1(\bx),\Phi_2(\by)}\approx
\sqrt{g}\frac{d\mG}{dR}
(R_{ij}R^{ij}(\bx)\delta(\bx-\by)-
\lambda R^2\delta(\bx-\by)- \nonumber
\\
&-&\nabla^i\nabla^j\delta(\bx-\by)\mG_{ijkl}R^{kl}(\bx)-
\nabla^i\delta(\bx-\by)\mG_{ijkl}\nabla^j
R_{kl}(\bx)-\nonumber \\
&-&\nabla^j\delta(\bx-\by)\mG_{ijkl}\nabla^i
R_{kl}(\bx)-\delta(\bx-\by)\mG_{ijkl}
\nabla^i\nabla^j R^{kl})+\nonumber \\
&+&\frac{1-\lambda}{D\lambda-1}
\nabla_k\nabla^k(
-R(\bx)\delta(\bx-\by)+(1-D)
\nabla_i\nabla^i\delta(\bx-\by))\equiv
\nonumber \\
&\equiv &
\mathbf{\triangle}(R,R_{ij},\bx,\by)+
\sqrt{g}\frac{d\mG}{dR}\frac{(1-\lambda)(1-D)}{D\lambda-1}
\nabla_i\nabla^i\nabla_j\nabla^j
\delta(\bx-\by) \nonumber \\
\end{eqnarray}
%
we find that the equation
(\ref{parttPhi2}) gives $v^2=0$.
In the same way the requirement of the
preservation of the constraint $\Phi_2$
implies
\begin{eqnarray}\label{partPhi2}
\partial_t\Phi_2\approx
\int d^D\bx (\pb{\Phi_2,\mH_T(\bx)} +
v^1(\bx)\pb{\Phi_2,\Phi_1(\bx)})=0 \ .
\nonumber \\
\end{eqnarray}
Using the fact that
$\pb{\Phi_2,\mH_T(\bx)}\neq 0$ and also
the equation (\ref{DBphi12}) we see
that (\ref{partPhi2}) can be solved for
$v^1$. In fact,  (\ref{DBphi12})
shows that
 $\Phi_1$ and
$\Phi_2$ are the second class
constraints and  previous analysis
that no additional
constraints have to be imposed on the
system.
 According to standard
analysis the second class
 constraints $\Phi_A,A=1,2$   have to
vanish strongly and allow us to express
two phase space variables as functions
of remaining physical phase space
variable. It is important to stress
that  the Poisson bracket between the
second class constraints depend on the
phase space variables so that it is
possible that it vanishes on some
subspace of phase space. On the other
hand from (\ref{DBphi12}) we see that
for $\lambda \neq 1$ this  Poisson
bracket is non-zero on the whole phase
space. On the other hand in case when
$\lambda=1$ we find that this Poisson
bracket vanishes for the subspace of
the phase space where $R_{ij}=0$. First
of all we have to check the consistency
of the condition  $R_{ij}=0$ with the
constraint $\mG(R)=0$. If yes, then we
see  that $\Phi_1$ is preserved during
the time evolution of the system and
hence it is not necessary to impose
additional constraint $\Phi_2\approx
0$. Moreover, the constraint
$\Phi_1\approx 0$ is replaced with the
set of more general constraints
$R_{ij}\approx 0$. These constraints
imply  that all metric components and
their conjugate momenta are
non-propagating degrees of freedom and
hence the theory on the subspace
$R_{ij}=0$ is effectively topological.
To conclude the Poisson bracket between
the constraints $\Phi_1,\Phi_2$ is
non-zero on the whole phase space for
$\lambda\neq 1$. In case of $\lambda=1$
it vanishes on the subspace $R_{ij}=0$
(on condition its consistency with
constraint $\mG(R)=0$) which is special
since it corresponds to effectively
topological theory.

Returning now to the constraints
$\Phi_1,\Phi_2$ we find that  it is
 very difficult to solve them in full generality.
\footnote{It is clear that the
linearized approximation gives the same
result as in \cite{Horava:2010zj} and
leads to the elimination of the scalar
graviton.} For that reason we restrict
to the general discussion of the
constraint structure of given theory
that allows us to determine the number
of physical degrees of freedom. To do
this we  note that there are $D(D+1)$
gravity phase space variables $g_{ij},
\pi^{ij}$, $2D$ variables $N_i,p^i$,
$2$ variables $\mA,p_\mA$. In summary
the total number of degrees of freedom
is
 $N_{D.o.f}=D^2+3D+2$.
 On the other
hand we have $D$  first class
constraints $\mH_i\approx 0$, $D$   first class
constraints  $p_i\approx 0 $, one   first class
constraint $p_\mA\approx 0$ and two
second class constraints $\Phi_1,
\Phi_2$. Then we have $N_{f.c.c}=2D+1$
  first class constraints and
 $N_{s.c.c.}=2$  second
class constraints. As a result  the
number of physical degrees of freedom
is \cite{Henneaux:1992ig}
\begin{equation}\label{pdf}
N_{D.o.f.}-2N_{f.c.c}-N_{s.c.c.}=
D^2-D-2 \
\end{equation}
that exactly corresponds to the number
of the phase space  physical degrees of freedom of $D+1$ dimensional
gravity.

We see that the phase space of the
Lagrange multiplier modified RFDiff HL
gravity  provides correct counting of
the  physical degrees of freedom of
gravitational theory. It is important
to stress that this is the theory
without global Hamiltonian constraint
and without additional $U(1)$ symmetry.
Further, this is the theory with
non-trivial symplectic structure.
To see this note that  the
Poisson bracket between the
constraints $\Phi_A$  can be written as
\begin{equation}
\pb{\Phi_A(\bx),\Phi_B(\by)}=
\triangle_{AB}(\bx,\by) \ ,
\end{equation}
where  the matrix $\triangle_{AB}$ has
following structure
\begin{equation}\label{triangleAB}
\triangle_{AB}(\bx,\by)=
\left(\begin{array}{cc}
0 & * \\
 \ * & * \\ \end{array}\right) \ ,
\end{equation}
where $*$ denotes non-zero elements.
It easy to see that matrix inverse
to (\ref{triangleAB}) has the form
\begin{equation}\label{triin}
(\triangle^{-1})^{AB}=
\left(\begin{array}{cc}
* & * \\
\ * & 0 \\ \end{array}\right) \ .
\end{equation}
  Now we observe
that
\begin{eqnarray}\label{gcon}
\pb{g_{ij}(\bx),\Phi_1(\by)}=0 \ ,
\pb{g_{ij}(\bx),\Phi_2(\by)}\neq 0 \ ,
\nonumber \\
\pb{\pi^{ij}(\bx),\Phi_1(\by)}\neq 0 \ ,
\pb{\pi^{ij}(\bx),\Phi_2(\by)}\neq 0 \ .
\nonumber \\
\end{eqnarray}
Then we find that the Dirac brackets
between canonical variables take the form
\begin{eqnarray}
& &\pb{g_{ij}(\bx),g_{kl}(\by)}_D=
-\int d\bz d\bz'
\pb{g_{ij}(\bx),\Phi_A(\bz)}
(\triangle^{-1})^{AB}(\bz,\bz')
\pb{\Phi_B(\bz'),g_{kl}(\by)}=0 \ ,
\nonumber \\
& &\pb{\pi^{ij}(\bx),\pi^{kl}(\by)}_D=
\nonumber \\
&=&-\int d\bz d\bz'
\pb{\pi^{ij}(\bx),\Phi_A(\bz)}
(\triangle^{-1})^{AB}(\bz,\bz')
\pb{\Phi_B(\bz'),\pi^{kl}(\by)}=
\Omega^{ijkl}(\bx,\by) \ ,
\nonumber \\
& &\pb{g_{ij}(\bx),\pi^{kl}(\by)}_D=
\pb{g_{ij}(\bx),\pi^{kl}(\by)}
-\nonumber \\
&-&\int d\bz d\bz'
\pb{g_{ij}(\bx),\Phi_A(\bz)}
(\triangle^{-1})^{AB}(\bz,\bz')
\pb{\Phi_B(\bz'),\pi^{kl}(\by)}=
\Omega_{ij}^{kl}(\bx,\by) \ ,
\nonumber \\
\end{eqnarray}
where the matrix $\Omega$ depends on
phase-space variables according to
(\ref{triin}) and (\ref{gcon}). The
fact that the symplectic matrix depends
explicitly on phase space variables
implies that it is nontrivial step to
proceed to the quantum mechanical
analysis of given system. In principle
it is possible to perform the abelian
conversion of Lagrange multiplier
modified RFDiff HL gravity to the
system with the first class constraints
following \cite{Batalin:1991jm}.
However the fact that the matrix
$\triangle_{AB}$ depends on the phase
space variables in non-trivial way we
can expect that the resulting
Hamiltonian and first class constrains
will contain infinite number of terms
and hence the analysis of given theory
will be very complicated.

 It is important to stress that the fact that
 the Hamiltonian is not given as linear
combination of constraints
 has an  important consequence for the
 stability of given theory. Explicitly,
 it is well known that some massive
 gravities are unstable since the
 Hamiltonian is not bounded from
 bellow. Alternatively, the instability
 of given theory is also indicated by
 presence of the ghosts (fields with
 wrong sign of kinetic term in the action)
  in the fluctuation spectrum. In case of the
 Lagrange multiplier modified
 RFDiff  HL gravity there is no such
a  ghost due to the fact that only
physical degrees of freedom propagate
(the scalar graviton is absent) and
hence linearized RFDiff invariant HL
gravity has positive definite
Hamiltonian. On the other hand  it is
not clear whether this holds  in
general case since  in order to fully
investigate the Hamiltonian of general
Lagrange multiplier modified HL gravity
 we should solve the
second class constraints and express
given  Hamiltonian in terms of physical
modes only. However as we argued above
this is very difficult task and hence
an  analysis of the  stability of
Lagrange multiplier modified HL gravity
  has not been performed
yet.


\section{More General Forms of
Lagrange Multiplier Modified HL
Gravities}\label{third} We would like
to stress that our work is based on the
formulation of the Lagrange multiplier
modified $F(R)$ gravities introduced in
 \cite{Capozziello:2010uv} where following
 form of
 Lagrange multiplier modified $F(R)$
gravity action was considered
\begin{equation}\label{SFF}
S=\int d^{D+1}x
\sqrt{-\hg}\left[F_1({}^{(D+1)}R)-
\Lambda \left(\frac{1}{2}\partial_\mu
{}^{(D+1)}R \hg^{\mu\nu}\partial_\nu
{}^{(D+1)}
R+F_2({}^{(D+1)}R)\right)\right] \ ,
\end{equation}
where $F_1$ and $F_2$ are arbitrary
functions and where ${}^{(D+1)}R$ is
$D+1$ dimensional scalar curvature and
where $\Lambda$ is Lagrange multiplier.
Note that the action (\ref{SFF}) is
invariant under full diffeomorphism of
 the target space-time. Introducing
two auxiliary fields $A,B$ we can
rewrite the action  (\ref{SFF}) into
the form
\begin{equation}\label{SFF2}
S=\int d^{D+1}x
\sqrt{-\hg}\left[F_1(A)- \Lambda
\left(\frac{1}{2}\partial_\mu A
\hg^{\mu\nu}\partial_\nu
A+F_2(A)\right)+
B({}^{(D+1)}R-A)\right]  \
\end{equation}
that is suitable for the generalization
to the case of $F(R)$ HL gravity.
Following
\cite{Carloni:2010nx,Chaichian:2010yi}
 we  find
the generalization of this action to the case
of HL gravity when we replace ${}^{(D+1)}R$ with
$\tilde{R}$ defined as
\begin{equation}
\label{tR}
 \tilde{R}= K_{ij}\mG^{ijkl}K_{kl}
+ \frac{2\mu}{\sqrt{-\hat{g}}} \partial_\mu \left(\sqrt{-\hat{g}}n^\mu K\right)
 -\frac{2\mu}{\sqrt{g}N} \partial_i \left(\sqrt{g}g^{ij}\partial_j N\right)
 -\mathcal{V}(g) \, ,
\end{equation}
where $\mu$ is constant,
$K=K_{ij}g^{ji}$. On the other hand
using the representation (\ref{SFF2})
we see that the Lagrange multiplier
modified $F(\tilde{R})$ HL gravity
is equivalent to the
 $F(\tilde{R})$ HL gravity
coupled to the  scalar field with
specific form of the action. The
similar situation
 was analyzed in
 \cite{Kluson:2010af}
in case of Lagrange
multiplier modified $F(R)$ gravities. It  was shown
 there
 that the Lagrange multiplier modification of the action
 implies specific Hamiltonian dynamics of the
 scalar field $A$ while the Hamiltonian
 structure of the part of the action
 corresponding to gravity degrees of freedom
 is the same as in case of original $F(R)$
 gravity. Clearly the
 same situation occurs in case of
 Lagrange multiplier modified $F(\tilde{R})$
 HL gravity. Even if such theory could be
 useful for further development of the
 cosmological models in the context of
 $F(\tilde{R})$ HL gravity it is also clear
 that given theory cannot solve the   scalar
 graviton problem that is general property
 of all projectable versions of HL
 gravities. However it is also clear
 that this scalar graviton can be
 eliminated when we extend
 $F(\tilde{R})$ HL action  with  the  term
 that is function of the scalar
 curvature and multiplied by  Lagrange
 multiplier.
 This procedure is completely the same
 as in
previous section so that we will not
repeat it here.

On the other hand we can consider more
general form of Lagrange multiplier
modified RFDiff HL gravity that is
inspired by the action (\ref{SFF}).
Explicitly, let us consider following
form of the Lagrange multiplier
modified RFDiff HL gravity
\begin{equation}
S=\frac{1}{\kappa^2} \int dt  d^D\bx
\sqrt{g}\left[\tK_{ij}\mG^{ijkl}
\tK_{kl}-\mV(g)+ \Lambda
\left(-\frac{1}{2}\tilde{\nabla}_n
R\tilde{\nabla}_n R +G\left(
\frac{1}{2}g^{ij}\partial_i R\partial_j
R\right)+F(R)\right)\right] \ ,
\end{equation}
where
\begin{equation}
\tilde{\nabla}_n=\partial_t
R-N^i\partial_i R \ ,
\end{equation}
and where $F$ and $G$ are general functions.
Introducing two auxiliary fields $A,\mA
$ we
can rewrite this action in an equivalent form
\begin{eqnarray}
S&=&\frac{1}{\kappa^2}
\int dt  d^D\bx
\sqrt{g}\left[
\tK_{ij}\mG^{ijkl}\tK_{kl}-\mV(g)+\mA(R-A)+\right. \nonumber \\
&+&\left.\Lambda
\left(-\frac{1}{2}\tilde{\nabla}_n
A\tilde{\nabla}_n A
+G\left(\frac{1}{2}g^{ij}\partial_i
A\partial_j A\right)+F(A)\right)\right]
\ .
\nonumber \\
\end{eqnarray}
The part of the action written on the
first line is the same as the action
(\ref{RFDaction}) when we identify
$\mG(R)$  with $(R-A)$. In order to see
whether given modification could be
useful it is instructive to perform the
Hamiltonian analysis of given action.
  As usual we begin
with the definition of conjugate momenta
\begin{eqnarray}
\pi^{ij}&=& \frac{1}{\kappa^2}\sqrt{g}\mG^{ijkl}\tK_{kl}
\ ,  \quad p^i
\approx 0 \ , \quad
p_\mA \approx 0   \ ,\nonumber \\
p_A&=&-\Lambda \sqrt{g}\tilde{\nabla}_n
A \ , \quad p_\Lambda\approx 0 \ .
\nonumber \\
\end{eqnarray}
From these relations we find
 $D+1$
primary constraints
\begin{equation}
p_i(\bx)\approx 0 \ ,  \quad
p_\mA(\bx)\approx 0 \ , \quad
p_\Lambda(\bx) \approx 0 \
\end{equation}
and the Hamiltonian
\begin{eqnarray}\label{HRFD2}
H&=&\int d^D\bx (\mH_T+N^i\mH_i+
v^\mA\Phi_\mA+v^\Lambda p_\Lambda+
v_i
p^ i)- \frac{1}{\kappa^2} \int d^D\bx
\sqrt{g} \mA (R-A) \ ,
\nonumber \\
\end{eqnarray}
where
\begin{eqnarray}\label{mHTAr}
\mH_T&=&\mH_T^{gr}+\mH_T^A \ , \quad
\mH_T^{gr}=
\frac{\kappa^2}{\sqrt{g}}\pi^{ij}\mG_{ijkl}\pi^{kl}
-\frac{1}{\kappa^2} \sqrt{g}\mV(g) \ ,
\nonumber \\
\mH_T^A&=&-\frac{\kappa^2}{2\Lambda\sqrt{g}}p_A^2-
\frac{1}{\kappa^2}\sqrt{g}\Lambda
\left(G\left(\frac{1}{2}g^{ij}\partial_i
A\partial_j A\right)+F(A)\right) \ ,
\quad
\nonumber \\
\mH_i&=&-2g_{il} \nabla_k \pi^{kl}+\partial_i A p_A \ . \nonumber \\
\end{eqnarray}
Now the time evolution of the primary
constraints implies
\begin{eqnarray}
\partial_t p_\mA &=&\pb{p_\mA,H}=
\frac{1}{\kappa^2} \sqrt{g}(R-A)\equiv
 \Phi_1 \approx 0 \ ,
\nonumber \\
\partial_t p_i&=&\pb{p_i,H}=
-\mH_i\approx 0
 \  ,  \nonumber \\
\partial_t p_\Lambda&=&
\pb{p_\Lambda,H}=-\frac{\kappa^2}{2\Lambda^2
\sqrt{g}}p_A^2+ \frac{1}{\kappa^2}
\sqrt{g}\left(G\left(\frac{1}{2}g^{ij}\partial_iA
\partial_j A\right)+F(A)\right)\equiv \Phi_2\approx
0 \  \nonumber \\
\end{eqnarray}
so that we have following secondary
constraints $\mH_i\approx 0 \ ,
\Phi_1\approx 0$ and $\Phi_2\approx 0$.
Including these constraints into
definition of the Hamiltonian we obtain
the total Hamiltonian in the form
\begin{equation}\label{Ht2}
H=\int d^D\bx (\mH_T+v^i p_i+v^\Lambda
p_\Lambda+v^\mA
p_\mA+v^1\Phi_1+v^2\Phi_2) +\bT_S(N^i)
\ ,
\end{equation}
where as usual we introduced the
smeared form of the diffeomorphism
constraint $\bT_S(N^i)=\int d^D\bx N^
i(\mH_i+p_\Lambda \partial_i\Lambda)$.

As the next step we analyze the
consistency of these secondary
constraints. Note that the constraint
$\bT_S(N^i)$ is preserved during the
time evolution of the system from the
same reason as in previous section.
Further, the  preservation of the
constraint $\Phi_1$ implies
\begin{eqnarray}\label{parttPhi1}
\partial_t\Phi_1=\pb{\Phi_1,H}=
-2\left(R_{ij}\pi^{ji}+\frac{\lambda}{D\lambda-1}
R\pi +\frac{(1-\lambda)}{\lambda D-1}
\nabla_k\nabla^k\pi\right)+\frac{p_A}{\Lambda}+
v^2\frac{p_A}{\Lambda^2 }=0 \ .
\nonumber \\
\end{eqnarray}
On the other hand the preservation  of
the constraint $p_\Lambda\approx 0$
implies
\begin{eqnarray}\label{parttplambda}
\partial_t p_\Lambda=
\pb{p_\Lambda,H}\approx
-v^2\frac{\kappa^2}{\Lambda^3\sqrt{g}}
p_A^2=0 \ .  \nonumber \\
\end{eqnarray}
If we combine this equation with the
equation  (\ref{parttPhi1}) we find an
additional constraint that has to be
imposed on the system
\begin{equation}\label{Phi1II}
\Phi_1^{II}
=R_{ij}\pi^{ji}+\frac{\lambda}{D\lambda-1}
R\pi +\frac{(1-\lambda)}{\lambda D-1}
\nabla_k\nabla^k\pi+\frac{p_A}{\Lambda}\approx
0 \
\end{equation}
while (\ref{parttplambda}) determines value
of the Lagrange multiplier $v_2=0$.

 It turns out that in
order to fully determine all Lagrange
multipliers we have to consider the
time evolution of the constraint
 $\Phi_1^{II}$ as well. Of course
  we also include
the expression $v^1_{II}\Phi_1^{II}$
into the definition of the total
Hamiltonian. Note also that we have
following non-zero Poisson brackets
\begin{equation}
\pb{\Phi_1^{II}(\bx),\mH_T(\by)}\ , \quad
\pb{\Phi_1^{II}(\bx),\Phi_1(\by)} \ ,
\quad
\pb{\Phi_1^{II}(\bx),\Phi_2(\by)} \ , \quad
 \pb{\Phi_1^{II}(\bx),
p_\Lambda(\by)} \ ,
\end{equation}
where the explicit form of these
Poisson brackets is not important for
us. However it is clear that the
presence of  the additional term in the
Hamiltonian has a consequence on the
time evolution of all constrains.
Explicitly
\begin{eqnarray}\label{partconfin}
\partial_t \Phi_1&=&\pb{\Phi_1,H}
\approx \int
d^D\bx\left(v^2(\bx)\pb{\Phi_1,\Phi_2(\bx)}+
v^1_{II}(\bx)
\pb{\Phi_1,\Phi^{II}_1(\bx)}\right)=0 \ ,
\nonumber \\
\partial_t\Phi_2&=&\pb{\Phi_2,H}
\approx \int d^D\bx \left(
\pb{\Phi_2,\mH_T(\bx)}+
v^1(\bx)\pb{\Phi_2,\Phi_1(\bx)}+\right.
\nonumber \\
&+& \left. v^2(\bx)\pb{\Phi_2,\Phi_2(\bx)}+
v^1_{II}(\bx)
\pb{\Phi_2,\Phi_1^{II}(\bx)}+
v^\Lambda(\bx)\pb{\Phi_2,p_\Lambda(\bx)}\right)\approx
0
 \nonumber \\
\partial_t\Phi_1^{II}&=&
\pb{\Phi_1^{II},H}\approx \int d^D \bx
\left(\pb{\Phi_1^{II},\mH_T(\bx)}+
v^1(\bx)\pb{\Phi_1^{II},\Phi_1(\bx)}+
\right.
\nonumber \\
&
&\left.+v^2(\bx)\pb{\Phi_1^{II},\Phi_2(\bx)}+
v^1_{II}(\bx)\pb{\Phi_1^{II},\Phi_1^{II}(\bx)}+
v^\Lambda
\pb{\Phi_1^{II},p_\Lambda(\bx)}\right)=0
\ ,
\nonumber \\
\partial_t p_\Lambda&=&
\pb{p_\Lambda,H}\approx \int
d^D\bx\left(v^2(\bx)
\pb{p_\Lambda,\Phi_2(\bx)}+
v^1_{II}(\bx)
\pb{p_\Lambda,\Phi_1^{II}(\bx)}\right)=0
\ .
\nonumber \\
\end{eqnarray}
We claim that these four equations can
be solved for four unknown
$v^1,v^2,v^1_{II}$ and $v^\Lambda$. In
fact, the last equation implies the
relation between $v^2$ and $v_1^{II}$
\begin{equation}
v_1^{II}= v^2\frac{\kappa^2}{\Lambda
\sqrt{g}}p_A \
\end{equation}
that together with the first  equation
in (\ref{partconfin}) implies
 $v^2=v^1_{II}=0$.
Then the second and third equations
simplify considerably
and  can be
solved for $v^1,v^\Lambda$ as functions
of canonical variables at least in
principle.
 The result of
this  analysis is that all Lagrange
multipliers are fixed. In other words
we found
 following four
second class constraints
\begin{equation}
\Phi_1(\bx)\approx 0 \ , \quad
\Phi_2(\bx)\approx 0 \ , \quad
\Phi_1^{II}(\bx)\approx 0  \ , \quad
p_\Lambda(\bx)\approx 0 \ .
\end{equation}
 Note that these constraints can be
explicitly solved on condition when we
replace Poisson brackets with Dirac
brackets.  From the last constraint we
find that $p_\Lambda(\bx)=0$. Further,
from $\Phi_1$ we find $A=R$ and then
from
 $\Phi_2$ we  express $\Lambda$ as
\begin{equation}
\Lambda^2=\frac{\kappa^4}{2g}
\frac{p_A^2}{(\frac{1}{2}g^{ij}
\partial_i R\partial_j R+F(R))} \ .
\end{equation}
Inserting this result into
$\Phi_1^{II}=0$ we find the  relation
between $g_{ij}$ and $\pi^{ij}$
\begin{equation}\label{Rijpij}
R_{ij}\pi^{ji}+\frac{\lambda}{D\lambda-1}
R\pi +\frac{(1-\lambda)}{\lambda D-1}
\nabla_k\nabla^k\pi+
\frac{\sqrt{2}}{\kappa^2}
\sqrt{g}\frac{1}{\sqrt{\frac{1}{2}g^{ij}
\partial_i R\partial_j R+F(R)}}=0 \ .
\end{equation}
Let us split the canonical momenta
$\pi^{ij}$ into  trace and traceless
parts as
\begin{equation}\label{pisplit}
\pi^{ij}=\tilde{\pi}^{ij}+\frac{1}{D}\pi
\ , \quad   g_{ij}\tilde{\pi}^{ji}=0 \
.
\end{equation}
Inserting (\ref{pisplit}) into
(\ref{Rijpij}) we can presume that it
can be solved for $\pi$ at least in
principle.
 Then the
reduced phase space is spanned by
$g_{ij},\tilde{\pi}^{ij}$ and $p_A$ so
that we have $D(D+1)$ physical degrees
of freedom.

Generally we can determine the
number of physical degrees as follows.
We have $D(D+1)$ metric phase space
variables $g_{ij},\pi^{ij}$, $2D$ phase
space variables $N_i,p^i$, $6$ phase
space variables
$\Lambda,p_\Lambda,A,p_A,\mA,p_\mA$. In
summary we have $N_{D. o.f.}=D^2+3D+6$
phase space degrees of freedom.
 On the other
hand we have $2D$ first class
constraints $\mH_i\approx 0 ,p^i\approx
0$, one first class constraint
$p_\mA\approx 0$ and $4$ second class
constraints $p_\Lambda\approx 0,
\Phi_1\approx 0, \Phi_2\approx 0$ and
$\Phi_1^{II} \approx 0$. In summary we
have $N_{f.c.c.}=2D+1$ first class
constraints and $N_{s.c.c.}=4$ second
class constraints. Then the number of
physical degrees of freedom is
\begin{equation}
N_{D.o.f.}-2 N_{f.c.c.}-N_{s.c.c.}=D^2-D=(D^2-D-2)+2
\end{equation}
where the expression in parenthesis
determines the number of physical
degrees of freedom of massless graviton
while the remaining part corresponds to the
scalar mode that is present in the theory.
 In other words the
generalized Lagrange multiplier
modified RFDiff HL gravity is not
sufficient for the elimination of the
scalar graviton and  should be only
considered as an interesting example of
the theory with reduced symmetry group.

 \noindent {\bf
Acknowledgements:}

 This work was also supported by the
Czech Ministry of Education under
Contract No. MSM 0021622409.

\end{document}